\documentclass[prd,aps,twocolumn,a4paper,floatfix]{revtex4-1}

\usepackage{graphicx,psfrag,mathrsfs}
\usepackage{slashed}
\usepackage{mathrsfs}
\usepackage{amsmath,amsfonts,amssymb,amsthm}
\usepackage{hyperref}
\usepackage{url}
\usepackage{comment,cancel}
\usepackage{accents}
\usepackage{ulem}

\makeatletter
\newcommand{\sbullet}{%
  \hbox{\fontfamily{lmr}\fontsize{.4\dimexpr(\f@size pt)}{0}
    \selectfont\textbullet}}
\DeclareRobustCommand{\mathbullet}{\accentset{\sbullet}}
\makeatother

\def\p{\partial}

\def\ul{\underline}
\def\non{\nonumber}
\def\mn{\mathring{\nabla}}
\def\mbn{\mathbullet{\nabla}}
\def\Gmmb{\Gamma[\mathring{\nabla},\mathbullet{\nabla}]}
\def\sD{\slashed{D}}

\newtheorem{thm}{Theorem}

\newtheorem{remark}{Remark}

\begin{document}

\title{High Order Asymptotic Expansions of a Good-Bad-Ugly Wave
  Equation}

\author{Miguel Duarte$^{1,2}$}
\author{Justin Feng$^2$}
\author{Edgar Gasper\'in$^{2,3}$}
\author{David Hilditch$^2$}

\affiliation{$^1$CAMGSD, Departamento de Matem\'atica, Instituto
  Superior T\'ecnico IST, Universidade de Lisboa UL, Avenida Rovisco Pais
  1, 1049 Lisboa, Portugal,
  $^2$CENTRA, Departamento de F\'isica, Instituto Superior
  T\'ecnico IST, Universidade de Lisboa UL, Avenida Rovisco Pais 1,
  1049 Lisboa, Portugal,
  \\$^3$ Institut  de  Math\'ematiques  de  Bourgogne  (IMB),  UMR  5584,
  CNRS,Universit\'e  de  Bourgogne  Franche-Comt\'e,
  F-21000  Dijon,  France}

\begin{abstract}
A heuristic method to find asymptotic solutions to a system of
non-linear wave equations near null infinity is proposed. The
non-linearities in this model, dubbed good-bad-ugly, are known to
mimic the ones present in the Einstein field equations (EFE) and we
expect to be able to exploit this method to derive an asymptotic
expansion for the metric in General Relativity (GR) close to null
infinity that goes beyond first order as performed by Lindblad and
Rodnianski for the leading asymptotics. For the good-bad-ugly model,
we derive formal expansions in which terms proportional to the
logarithm of the radial coordinate appear at every order in the bad
field, from the second order onward in the ugly field but never in the
good field. The model is generalized to wave operators built from an
asymptotically flat metric and it is shown that it admits
polyhomogeneous asymptotic solutions. Finally we define stratified
null forms, a generalization of standard null forms, which capture the
behavior of different types of field, and demonstrate that the
addition of such terms to the original system bears no qualitative
influence on the type of asymptotic solutions found.
\end{abstract}

\maketitle

\section{Introduction}

Although the concept of null infinity has been present in General
Relativity (GR) for a long time, dating back at least to the works of
Penrose, Newman, Bondi and Sachs ---see for
instance~\cite{Pen63,NewPen62, Bon62, Sac62, Sac62a}, there are still
questions to be answered about the structure of spacetime nearby. From
the point of view of mathematical relativity, null infinity plays a
central role in the resolution of open problems such as the weak
cosmic censorship conjecture, the (non) peeling properties and global
stability analysis of spacetimes.  From an astrophysical perspective,
null infinity should also play an important role since gravitational
radiation is not localizable and hence it is only well defined at null
infinity. The latest achievements in gravitational wave astronomy are
coupled to advances in numerical relativity since the former rely on
the accurate calculation of waveforms from astrophysically relevant
scenarios. However, the waveforms that are routinely computed in
numerical relativity codes are evaluated at a large but finite radius
and extrapolated to infinity. Despite that the latter has proven to
work, from a mathematical point of view these wave forms should thus
be computed directly at null infinity. There have been various
approaches to include null infinity in the computational domain. For
instance, the work of H\"ubner~\cite{Hub99,Hub01} and
Frauendiener~\cite{DouFra16} makes use of the conformal Einstein field
equations (CEFE) introduced by Friedrich~\cite{Fri81,Fri81a} building
upon Penrose's idea of bringing null infinity to a finite coordinate
distance by means of a conformal
compactification~\cite{Pen63}. Although the CEFE provide a geometric
approach to the problem of the inclusion of null infinity, the
standard methods of numerical relativity that have proven to work well
for the strong field region of spacetimes of physical interest cannot
be {\it trivially} lifted over. In particular, this approach has not
yet been used for compact binary evolutions.

To overcome this situation a variety of approaches have been pursued,
including using Cauchy-Characteristic Matching~\cite{Win12} and the
use of a suitable hyperboloidal initial value problem. The latter
involves providing initial data on a hyperboloidal slice, a spacelike
hypersurface that intersects future null infinity. These slices are
not Cauchy hypersurfaces, as their domain of dependence does not cover
the whole spacetime. The main technical problem with this approach is
that it results in formally singular terms which complicates
significantly their mathematical analysis as well as their numerical
implementation.  Although challenging, this type of singular equation
has been treated numerically in spherical
symmetry~\cite{Zen07,VanHusHil14,VanHus14,Van15}.

In light of the above, a relevant problem to be solved on the
analytical side, with direct implications for numerical work, is the
construction of an alternative to the CEFEs using more standard
formulations of the Einstein field equations (EFE) but insisting on
including null infinity. A recent proposal to make inroads into the
construction of such a formulation in the hyperboloidal set up is to
use a dual frame approach~\cite{Hil15}, which consists essentially of
decoupling coordinates from the tensor basis and carefully choosing
each of them. This allows one to write the EFE in generalized harmonic
gauge (GHG) and then solve them in hyperboloidal coordinates. Various
different aspects of this proposal have been
investigated~\cite{HilHarBug16,GasHil18,GasGauHil19,GauVanHil21}. Here
we give just a brief overview. An essential prerequisite for this to
work is the satisfaction of the coordinate \textit{lightspeed
  condition}~\cite{GasHil18}. As discussed in~\cite{GasHil18}, the
coordinate lightspeed condition is related to the \textit{weak null
  condition}~\cite{LinRod03}.  The former is the requirement that
derivatives of the radial coordinate lightspeeds have a certain
fall-off near null infinity, while the latter is expected to be a
sufficient condition on the non-linearities of a quasilinear wave
equation for establishing small data global existence. Although it has
not been shown in full generality that the weak null condition implies
small data global existence, a recent work by Keir~\cite{Kei17} proved
that if a system of quasilinear wave equations satisfies the
\textit{hierarchical weak null condition}, then small data global
existence is guaranteed. H\"ormander's \textit{asymptotic system}, a
heuristic method that predicts the fall-off of solutions to systems of
quasi-linear wave equations, was used in~\cite{GasHil18} to show that
through constraint addition, one can guarantee that the resulting
field equations satisfy the lightspeed condition beyond the initial
data.

The work in~\cite{HilHarBug16} and~\cite{GasHil18} together shows that
formally singular terms can be avoided by using the dual foliation
formalism~\cite{Hil15} in combination with hyperboloidal coordinates
and GR in GHG. However, even the simplest choice of variables shows
the existence of metric components with a fall-off of the
type~$O(R^{-1}\log R)$, with~$R$ a suitably defined radial
coordinate. This can cause problems in numerical
evolutions. In~\cite{GasGauHil19}, the authors use a toy model
composed of wave equations with non-linearities of the same kind as
those present in the EFE to show that these logarithmically divergent
terms can be explicitly regularized by a non-linear change of
variables. This toy model is called the \textit{good-bad-ugly} model
as it splits the evolved fields into three categories according to
their fall-off near null infinity, and it is known to satisfy the weak
null condition.

In this work we generalize earlier results on the
\textit{good-bad-ugly} model, laying out a heuristic method to predict
the type of decay of terms beyond the leading ones in a very large
class of systems of non-linear wave equations near null infinity.
This provides us with the knowledge of where log-terms may appear in
asymptotic expansions so that we are able to manage those terms
appropriately in the numerics. The adjective heuristic in this context
is used to emphasize that the connection between the original weak
null condition introduced by Lindblad and Rodnianski
in~\cite{LinRod03} and small data global existence has not been proven
yet, at least not in full generality. To remove this adjective one
would need to prove suitable estimates for our formal expansions. This
goes beyond the scope of the present article.

In Sections~\ref{setup} and~\ref{assumptions} we outline our geometric
setup and basic assumptions. In section~\ref{section:flatGBU} we begin
our analysis proper by considering the same model used
in~\cite{GasGauHil19}. We show by induction that the bad field may
have logs at every order in~$R^{-1}$ and the ugly field may have logs
from second order onward under certain conditions, while the good
field must have no logs at all. Additionally, we present a recursion
relation that gives the coefficient associated with any power
of~$R^{-1}$ of the evolved fields in terms of the previous order,
ultimately in terms of the initial data. In
section~\ref{section:gencase} the model is generalized to allow the
wave operator to be built from a general asymptotically flat metric
whose components are allowed to depend analytically on the evolved
fields. An induction proof analogous to the one shown in
section~\ref{section:flatGBU} is presented to assert that with such a
wave operator, the equations are much more deeply coupled and hence
all fields may inherit logs from one another, the main difference
being the order at which they are allowed to first appear. In fact,
the proof shows that the \textit{good-bad-ugly} model allows for
asymptotic solutions which can be written as polyhomogeneous
expansions, loosely speaking inverse power-law decay in~$R$ but with
logarithmic obstructions, close to null infinity. Finally, in
section~\ref{section:nullforms} we generalize the model even further
by adding arbitrary linear combinations of what we call stratified
null forms, a generalization of the standard notion that knows about
the behavior of the three different types of field, to the original
system and showing that these terms do not affect the the
proof. Because we keep the metric general, naturally it is not
possible to find a final recursion relation for the evolved variables
as is done in section~\ref{section:flatGBU}, but once the exact
dependence of the metric on the evolved fields is given, it should be
possible to find such a relation. Concluding remarks are given in
section~\ref{conclusion}.

\section{Geometric Set up}\label{setup}

\paragraph*{Representation of the metric:} Latin indices will be
used as abstract tensor indices while Greek indices will be used to
denote spacetime coordinate indices. We assume the existence of a
Lorentzian metric~$g_{ab}$ with Levi-Civita connection~$\nabla$ and
introduce the coordinate system~$X^{\ul{\alpha}}=(T,X^{\ul{i}})$,
which we require to be asymptotically Cartesian in a sense clarified
below. We raise and lower indices with the spacetime metric~$g_{ab}$
exclusively. Let~$\p_{\ul{\alpha}}$ and~$dX^{\ul{\alpha}}$ be the
corresponding vector and co-vector bases. The covariant derivative
associated to~$X^{\ul{\alpha}}$ is~$\mn$ and its Christoffel symbols
are defined by,
\begin{align}
	\Gamma[\mn]_{a}{}^{b}{}_{c} =
        (\mn_a\p_{\ul{\alpha}}^b)(dX^{\ul{\alpha}})_c\,.
\end{align}
Additionally we define shell
coordinates~$X^{\ul{\alpha}'}=(T',X^{\ul{i}'})=(T,R,\theta^A)$, where
the radial coordinate $R$ is related to $X^{\ul{i}}$ in the usual
manner
as~$R^2=(X^{\ul{1}})^2+(X^{\ul{2}})^2+(X^{\ul{3}})^2$. Let~$\p_{\ul{\alpha}'}$
and~$dX^{\ul{\alpha}'}$ be the corresponding vector and co-vector
bases. Shell coordinates have an associated covariant
derivative~$\mbn$ with Christoffel symbols,
\begin{align}
  \Gamma[\mbn]_{b}{}^{a}{}_{c}
  = (\mbn_b\p_{\ul{\alpha}'}^a)(dX^{\ul{\alpha}'})_c\,.
\end{align}
The transition tensor between the two covariant derivatives is defined
by,
\begin{align}
\Gmmb_{a}{}^{b}{}_{c}v^c = \mn_av^b - \mbn_av^b\,,
\end{align}
where~$v^a$ is an arbitrary vector field. We define outgoing and
incoming null vectors according to,
\begin{align}\label{eq:psiinshellchart}
  \psi^a&=\p_T^a+\mathcal{C}_+^R\p_R^a\,,\nonumber\\
  \ul{\psi}^a&=\p_T^a+\mathcal{C}_-^R\p_R^a\,,
\end{align}
where~$\mathcal{C}_+^R$ and~$\mathcal{C}_-^R$ are fixed by the
requirement that~$\psi^a$ and~$\ul{\psi}^a$ are null vectors with
respect to the metric~$g_{ab}$. We furthermore define two null
co-vectors by,
\begin{align}\label{eq:xitoeta}
\sigma_a&=e^{-\varphi}\psi_a\,,\quad \ul{\sigma}_a=e^{-\varphi}\ul{\psi}_a\,,
\end{align}
where~$\varphi$ is fixed by requiring that
\begin{align}
	\sigma_a\p_R^a=-\ul{\sigma}_a\p_R^a=1\,,
\end{align}
so we can write,
\begin{align}\label{eq:xiinshellchart}
  \sigma_a&=-\mathcal{C}_+^R\nabla_aT+\nabla_aR
  + \mathcal{C}^+_A\nabla_a\theta^A\,,\nonumber\\
  \ul{\sigma}_a&=\mathcal{C}_-^R\nabla_aT-\nabla_aR
  + \mathcal{C}^-_A\nabla_a\theta^A\,.
\end{align}
We choose to write the inverse spacetime metric as,
\begin{align}\label{eq:metricrepresentation}
g^{ab}=-2\tau^{-1}e^{-\varphi}\,\psi^{(a}\ul{\psi}^{b)}+\slashed{g}^{ab} \,,
\end{align}
where the null vectors satisfy,
\begin{align}
	&\sigma_a\psi^a=\ul{\sigma}_a\ul{\psi}^a=0\,,\non\\
	&\sigma_a\ul{\psi}^a=\ul{\sigma}_a\psi^a=-\tau\,,
\end{align}
with~$\tau:=\mathcal{C}_+^R-\mathcal{C}_-^R$.
In~\eqref{eq:metricrepresentation}, the normalization of the first
term is carefully chosen so that,
\begin{align}
	\slashed{g}^{ab}\sigma_b=\slashed{g}^{ab}\ul{\sigma}_b=0\,,
\end{align}
and~$\slashed{g}^b{}_a$ therefore serves as a projection operator
orthogonal to these two covectors. Note that~$\slashed{g}^{ab}$ is not
the inverse induced metric on level sets of~$T$ and~$R$, as it is not
orthogonal to~$\nabla_aT$ or~$\nabla_aR$, but rather to~$\sigma_a$
and~$\ul{\sigma}_a$. At first sight this seems unsatisfactory
geometrically, but since we will be heavily using the method of
characteristics it turns out that to be much more convenient to have a
simple representation of the vectors~$\psi^a$ and~$\ul{\psi}^a$ than
the covectors~$\sigma_a$ and~$\ul{\sigma}_a$. Our convention
for~$\slashed{g}^{ab}$ follows from this fact. The metric can be
written naturally as,
\begin{align}
  g_{ab}=-2\tau^{-1}e^{\varphi}\,\sigma_{(a}\ul{\sigma}_{b)}+\slashed{g}_{ab}\,.
\end{align}
This way we have ten independent metric functions, namely,
\begin{align}
  &\mathcal{C}_\pm^R\,,\,\mathcal{C}_A^\pm \,,\,
  \varphi\,,\,\slashed{g}^{ab}\,.
\end{align}
Tensors projected with~$\slashed{g}_a{}^b$ will be denoted adding a
slash to the kernel letter~$\slashed{T}_{ab} \equiv
\slashed{g}_a{}^c\slashed{g}_{b}{}^dT'_{cd}$. The covariant derivative
associated to~$\slashed{g}_{ab}$ will be denoted as~$\sD$, so that,
\begin{align}
\sD_b v^a := \slashed{g}_{c}{}^{a}\slashed{g}_{b}{}^{d}\nabla_d v^c\,,
\end{align}
where the vector satisfies~$v^a=\slashed{g}_{b}{}^{a}v^b$. We take the
obvious extension for higher rank tensors. Similarly, we define the
covariant derivative~$\mathring{\sD}$ as,
\begin{align}
\mathring{\sD}_b v^a := \slashed{g}_{c}{}^{a}\slashed{g}_{b}{}^{d}\mn_d v^c\,.
\end{align}
We define the vector field~$T^a:=\p_T^a$ and denote the covariant
derivative in the direction of~$T^a$ as~$\nabla_T$. Analogous notation
will be used for directional derivatives along other vector
fields. Because we will have to deal with terms proportional
to~$R^{-n}(\log R)^m$, the use of the term `order' might be confusing,
as for the same power of~$R^{-1}$ different values of~$m$ give rise to
different decays. To clarify that, throughout this work, `order~$n$'
will denote terms proportional to~$R^{-n}$.

\paragraph*{The \textit{good-bad-ugly} system:} We introduce the
following model,
\begin{align}
        \mathring{\square} g &= 0\,,\non\\
        \mathring{\square} b &= (\nabla_T g)^2\,,\non\\
        \mathring{\square} u &= \tfrac{2}{R}\nabla_T u\,,
        \label{gbu}
\end{align}
where~$g$, $b$ and~$u$ stand for~{\it good, bad} and {\it ugly}
fields, respectively, $\mathring{\square}$ is called the reduced wave
operator and it is defined by~$g^{ab}\mn_a\mn_b$. Because we will only
be concerned with the large~$R$ regime, we are not concerned with
regularity at the origin. Therefore, for simplicity, we have adjusted
the final equation of the model given in~\cite{GasGauHil19} so that
the source term appears with a simple coefficient~$2/R$. The
metric~$g^{ab}$ can be taken to depend on the evolved fields
themselves in a manner we will expand upon below. The leading order of
the particular case where the metric~$g^{ab}$ is the Minkowski metric
was studied in detail in~\cite{GasGauHil19}.

\section{Assumptions}\label{assumptions}

We need to place certain assumptions on the evolved fields and metric
functions that will allow us to formally equate terms of the same
order in~\eqref{gbu} and retrieve simpler equations that are
satisfied, order-by-order, by~$g$, $b$ and~$u$ close to null infinity.

\paragraph*{Evolved fields:} We define a null
tetrad~$\{\psi,\ul{\psi},X_1,X_2\}$, where~$X_1$ and~$X_2$ are
orthogonal to~$\psi^a$ and~$\ul{\psi}^a$ and normalized respect
to~$g_{ab}$, namely, $g_{ab}X_A^a\psi^b=g_{ab}X^a_A\ul{\psi}^b=0$
and~$g_{ab}X_A^aX_B^b=\delta_{AB}$, with $A=1,2$. Let~$\omega_{g,b}$
be any field in~$\{g,b\}$ or any first derivative thereof. Based on
insight from~\cite{Kei17} we will assume first derivatives of~$\omega$
to have the following behavior near null infinity,
\begin{align}\label{weaknull1}
\omega_{g,b} = o^+(R^{-n})\Rightarrow
    \begin{cases}
        \nabla_{\psi}\omega_{g,b} = o^+(R^{-n-1})\\
	    \nabla_{\ul{\psi}}\omega_{g,b} = o^+(R^{-n})\\
        \nabla_{X_A}\omega_{g,b} = o^+(R^{-n-1})
    \end{cases}
    \,,
\end{align}
with~$A\in\{1,2\}$. Here,~$f=o^+(h)$ as~$R\rightarrow\infty$ is
defined as the condition,
\begin{align}\label{deflittleo}
  \exists \epsilon>0 :
  \lim_{R\rightarrow\infty} \frac{f}{hR^{-\epsilon}}=0\,.
\end{align}
Note that this condition is a more restrictive version of~$f=o(h)$,
which can be informally stated as \textit{$f$ falls-off faster
  than~$h^{1+\epsilon}$ as $R$ goes to infinity}. In
particular,~$o^+(h)=o(hR^{-\epsilon})$. The reason why we make this
slightly stronger assumption will become apparent once we start
integrating error terms in the next section. Derivatives
along~$\psi^a$ and~$X_A$ are called good derivatives, while the ones
along~$\ul{\psi}^a$ are called bad derivatives. This naming convention
is motivated by the fact that, for fields satisfying equations like
ours, the former improve the fall-off of the argument, whereas the
latter do not. Let~$\omega_u$ be the field~$u$ or any first derivative
thereof. We know from~\cite{GasGauHil19} that, in the case
that~$\mathring{\square}$ is built from the Minkowski metric,
derivatives of~$u$ have different asymptotics from the other fields,
so we assume,
\begin{align}\label{weaknull2}
\omega_u = o^+(R^{-n})\Rightarrow
    \begin{cases}
        \nabla_{\psi}\omega_u = o^+(R^{-n-1})\\
	    \nabla_{\ul{\psi}}\omega_u = o^+(R^{-n-1})\\
        \nabla_{X_A}\omega_u = o^+(R^{-n-1})
    \end{cases}
    \,.
\end{align}
We make this set of assumptions using for example~$o^+(R^{-n})$
instead of the more restrictive~$O(R^{-n-1})$, because previous work
has shown that similar equations have asymptotic solutions
proportional to, for instance,~$R^{-1}\log R$, and naively using
big~$O$ notation would not permit such
solutions~\cite{GasGauHil19}. Furthermore, we are interested in
physically relevant solutions, so we cannot allow fields which do not
decay near null infinity. Therefore, we restrict our attention to a
space of initial data in which there is decay near null infinity,
i.e.,
\begin{align}
	g=o^+(1),\quad b=o^+(1), \quad u=o^+(1)\,.
\end{align}
Let~$\mathcal{S}$ be a Cauchy surface defined by~$T=T_0$,
where~$T_0$ is a constant. In order to allow for nonzero ADM mass and
linear momentum, we choose initial data which decays at spacelike
infinity as,
\begin{align}
&\phi = O_{\mathcal{S}}(R^{-1})\,,\non\\
&\nabla_T \phi = O_{\mathcal{S}}(R^{-2})\,,\label{assumptionTimeder}
\end{align}
where the subscript~$\mathcal{S}$ is used to say that this fall-off is
required on a spatial slice, rather than at null infinity. This is not
the most general choice of initial data which allows for nontrivial
ADM mass and linear momentum, but it is broad enough to include
practically all spacetimes of interest.

\paragraph*{Metric functions:} As derivatives of our metric functions
will, in general, be present in the field equations, we must have a
way to collect them in powers of $R^{-1}$. Therefore we require that
the metric functions may be written as,
\begin{align}
&\mathcal{C}_\pm^R = \pm1 + \gamma^\pm_1\,,\non\\
&\mathcal{C}_A^\pm = R\gamma_2^\pm\,, \non \\
&\varphi = \gamma_3\,, \non \\
&\slashed{g}^{ab} = \slashed{\eta}^{ab} + R^{-2}\gamma_4{}^{ab}\,,
\label{asympflatness}
\end{align}
where~$\slashed{\eta}^{ab}$ is the inverse metric on the round
2-sphere of radius~$R$ and the~$\gamma$'s are analytic functions of
only the evolved fields in a neighborhood of null
infinity,~$\gamma=\gamma(g,b,u)$. Moreover, we are interested in
studying spacetimes with metrics that asymptote to the Minkowski
metric as we approach~$\mathscr{I}^+$, i.e. asymptotically flat
metrics. This implies that the~$\gamma$ functions must go to zero as
we approach null infinity, that is,
\begin{align}
	\gamma(g,b,u)|_{\mathscr{I}^+}=0\,.
\end{align}

\section{Flat metric}\label{section:flatGBU}

In this section we will study the asymptotics of the
\textit{good-bad-ugly} system~\eqref{gbu} with a flat metric
$g^{ab}=\eta^{ab}$ near null infinity, where $\eta^{ab}$ is the
inverse Minkowski metric. Based on our assumptions on the decay of the
fields and their derivatives~\eqref{weaknull1} and~\eqref{weaknull2},
we will equate terms of the same order to find simpler equations that
will reveal the asymptotics of~$g$, $b$ and~$u$. Moreover, this
section will serve as a toy model for the next, where we analyze the
system for a general asymptotically flat metric. Requiring that the
metric be flat implies,
\begin{align}
	&\mathcal{C}_\pm^R = \pm1\,,\non\\
	&\varphi=\mathcal{C}_A^\pm = 0\,, \non\\
	&\slashed{g}^{ab} = \slashed{\eta}^{ab}\,,\label{flatmetricfuncs}
\end{align}
meaning that all metric components are given, the only unknowns being
the evolved fields themselves. The inverse Minkowski metric can then
be written in terms of null vectors in the following way,
\begin{align}\label{eq:metricrepresentationflat}
  \eta^{ab}=-\psi^{(a}\ul{\psi}^{b)}+\slashed{\eta}^{ab} \,,
\end{align}
where $\psi^a$ and $\ul{\psi}^a$ reduce to,
\begin{align}
  &\psi^a=\p_T^a+\p_R^a\,,\non\\
  &\ul{\psi}^a=\p_T^a-\p_R^a\,.\label{eq:xiinshellchartflat}
\end{align}
The method we want to implement relies on integrating the equations we
get along different vector fields. These integrations are made simpler
if we work under a Bondi-like approach, rewriting the incoming null
vector~$\ul{\psi}^a$ as a function of the timelike vector~$\p_T^a$ and
the outgoing null vector~$\psi^a$,
\begin{align}
	\ul{\psi}^a = 2\p_T^a - \psi^a\,.
\end{align}
Let~$\phi$ be any field in~$\{g,b,u\}$. We can expand the wave
operator in the following way,
\begin{align}\label{toybox1}
	\mathring{\square} \phi = \left(-2\nabla_{\psi}\nabla_T -
        \frac{2}{R}\nabla_T + \frac{2}{R}\nabla_{\psi} +
        \nabla_{\psi}^2+ \cancel{\Delta}\right)\phi\,,
\end{align}
where,
\begin{align}
   \cancel{\Delta}\phi := \mathring{\sD}^a\mathring{\sD}_a\phi\,,
\end{align}
is the Laplace operator on the $2$-sphere of radius~$R$. We treat each
equation in~\eqref{gbu} separately, and starting with the first.

\subsection{The Good field}

\begin{figure}[htbp]
\includegraphics[width=0.45\textwidth]{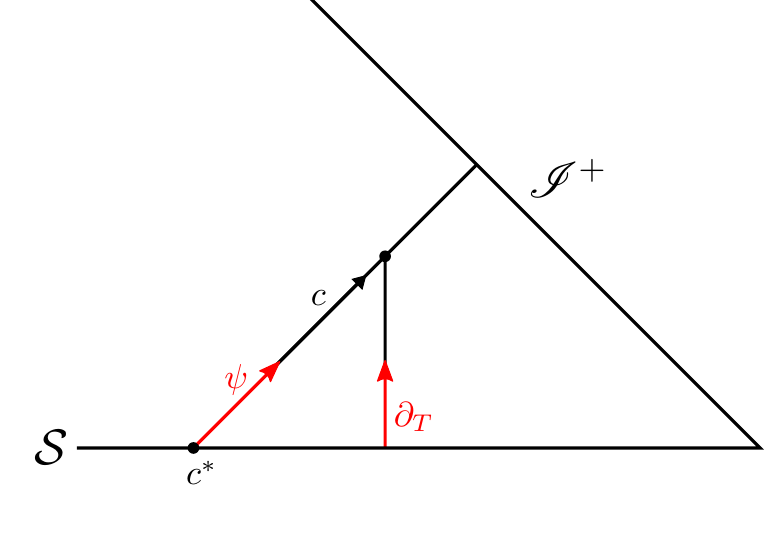}
\caption{A schematic of our geometric setup. The method proceeds first
  by integrating out along~$c$, an integral curve of the outgoing
  null-vector~$\psi^a$ and then up along integral curves of~$\p_T^a$.
  \label{fig:Int_Fig}}
\end{figure}

\paragraph*{Motivation for induction hypothesis:} We begin by
rescaling~$g$ as,
\begin{align}\label{gresc}
	\mathcal{G}_1 := g R\,,
\end{align}
and plugging that into~\eqref{toybox1} to get the equation,
\begin{align}\label{gmot1}
  -2\nabla_\psi\nabla_T\mathcal{G}_1 + \nabla_\psi^2\mathcal{G}_1
  + \cancel{\Delta}\mathcal{G}_1 = 0\,.
\end{align}
It is known from~\cite{Kei17} that~$g$ and its derivatives
satisfy~\eqref{weaknull1}, so there is a hierarchy among the different
terms in~\eqref{gmot1} which allows us to neglect some of them and end
up with a simpler equation that determines~$\mathcal{G}_1$ to leading
order. This is just H\"ormander's first order asymptotic system for
the wave equation. We know that the second and third terms
in~\eqref{gmot1} are of higher order than the first one because they
contain two good derivatives, so the first term must vanish by itself
to leading order,
\begin{align}\label{gmot2}
\nabla_\psi\nabla_T\mathcal{G}_1 = o^+(R^{-1})\,.
\end{align}
We want to integrate this expression along integral curves of~$\psi^a$
and then~$\p_T^a$. Since integrating error terms is not completely
straightforward, we dedicate a paragraph after the present one to
proving that error terms remain suitably small under integration. For
now we will only pay attention to the leading
contributions. Let~$c(s)$ be an integral curve of the vector
field~$\psi^a$ that passes through the point~$c^{*} \in \mathcal{S}$
at a fiduciary value of~$s=s^{*}$ and integrate equation~\eqref{gmot2}
along that curve to get,
\begin{align}\label{goodfirstint1}
	\nabla_T\mathcal{G}_1 \simeq \dot{g}_1(c^*) \,,
\end{align}
where~$\dot{g}_1(c^*) = \nabla_T g_1|_{c(s^*)}$. We use~$\simeq$ to
represent equality at large radius up to error terms that decay faster
than those displayed in the expression. For example, we can
write~$f\simeq R^{-1}$ as short-hand for~$f=R^{-1}+o^+(R^{-1})$. As we
have not specified the curve along which the integration was
performed,~\eqref{goodfirstint1} is valid for any~$c$ and so we can
write ($\psi^*$ denoting the dependence on the initial data at $c^*$),
\begin{align}\label{nablaTG}
	\nabla_T\mathcal{G}_1 \simeq \dot{g}_1(\psi^*) \,,
\end{align}
where~$\dot{g}_1(\psi^*)$ is fixed along any particular integral curve
of~$\psi^a$. The exact same method will be used in the rest of this work
whenever integrating along integral curves
of~$\psi^a$. Integrating~\eqref{nablaTG} in~$T$ we get,
\begin{align}\label{G}
\mathcal{G}_1 \simeq \int_{T_0}^T\dot{g}_1(\psi^*)dT' + m_{g,1}\,,
\end{align}
where~$m_{g,1}$ is a scalar function that is independent of~$T$ and we
choose it to be independent of $R$ as well. In fact, our choice of
initial data~\eqref{assumptionTimeder} requires that all of
the~$m_{\phi,1}$ functions throughout the rest of this work are
independent of~$R$. Moreover, because we
impose~\eqref{assumptionTimeder}, that choice implies
that~$\dot{g}_1(\psi^*)$ fall-off like~$O_{\mathcal{S}}(R^{-1})$
because,
\begin{align}
	\nabla_T g \simeq \frac{\dot{g}_1(\psi^*)}{R}\,.
\end{align}
Let us now analyze the leading error terms. We define the
function~$\mathcal{G}_2$ as,
\begin{align}\label{G'/R}
	\frac{\mathcal{G}_2}{R} := \mathcal{G}_1 - G_1(\psi^*)=o^+(1)\,,
\end{align}
where,
\begin{align}\label{G1}
	G_1(\psi^*) := \int_{T_0}^T\dot{g}_1(\psi^*)dT' + m_{g,1}\,,
\end{align}
and assume that it also satisfies~\eqref{weaknull1}. Then
from~\eqref{gmot1} we get,
\begin{align}\label{gmot3}
  2\nabla_\psi\left(\frac{1}{R}\nabla_T\mathcal{G}_2\right)
  + \nabla_\psi^2\frac{\mathcal{G}_2}{R}
  + \cancel{\Delta}
  \left(G_1 + \frac{\mathcal{G}_2}{R}\right) = 0\,.
\end{align}
Here it pays off to introduce the operator,
\begin{align}
	\tilde{\cancel{\Delta}}:=R^2\cancel{\Delta}\,,
\end{align}
which makes the order in~$R^{-1}$ explicit, as~$\cancel{\Delta}$
amounts to two good derivatives. Once again, collecting the lowest
order terms we get, asymptotically,
\begin{align}\label{gmot4}
  2\nabla_\psi\left(\frac{1}{R}\nabla_T\mathcal{G}_2\right)
  +\frac{1}{R^2} \tilde{\cancel{\Delta}}G_1 = o^+(R^{-2})\,.
\end{align}
which gives,
\begin{align}\label{G'2}
  \mathcal{G}_2 \simeq -\frac{1}{2}\int_{T_0}^
  T\tilde{\cancel{\Delta}}G_1dT'
  + R\int_{T_0}^T\dot{g}_2(\psi^*)dT' + m_{g,2}\,.
\end{align}
In order to integrate along an integral curve of~$\psi^a$ we
parameterize~$c$ using the radial coordinate~$R$ so that we get,
\begin{align}
\int_{c} \frac{1}{R^n}dR = -\frac{1}{(n-1)R^{n-1}}\,,
\end{align}
along any~$c$.
Note that the second term on the RHS of~\eqref{G'2}
grows like~$R$, which would contradict~\eqref{G'/R}. However, if we
were to write~$\nabla_Tg$ with what we know already at first order
and~\eqref{G'2} as it is, we would get,
\begin{align}
  \nabla_Tg& = \frac{\nabla_TG_1}{R}
  +\frac{\nabla_T\mathcal{G}_2}{R^2}\non\\
  &\simeq\frac{1}{R}(\dot{g}_1+\dot{g}_2)
  +\frac{1}{2}\tilde{\cancel{\Delta}}G_1\,,\label{g2=0}
\end{align}
which implies that~$\dot{g}_2$ can be absorbed into
$\dot{g}_1$. Therefore we can choose solutions with~$\dot{g}_2=0$
without any loss of generality, so that,
\begin{align}\label{G'3}
	\mathcal{G}_2 \simeq G_2(\psi^*)\,,
\end{align}
where,
\begin{align}
  G_2(\psi^*) := \frac{1}{2}\int_{T_0}^T\tilde{\cancel{\Delta}}G_1dT'
  + m_{g,2}\,.
\end{align}
Equations~\eqref{G} and~\eqref{G'3} suggest that the field~$g$ may be
written as,
\begin{align}\label{gn}
	g = \sum_{n=1}^\infty \frac{G_n(\psi^*)}{R^n}\,,
\end{align}
and we prove that result shortly. However it is worth pausing here for
a moment to take a closer look at how the error terms behave under
integration.

\paragraph*{Integration of error terms:} What we aim to show here is
that if~$f=o^+(R^{-n})$ then,
\begin{align}\label{interror3}
	\int fdR = o^+(R^{-n+1})\,,
\end{align}
for all~$n\in \mathbb{N}$. In other words, we want to show,
\begin{align}\label{interror1}
	\lim_{R\rightarrow \infty}R^{n-1+\epsilon}\int fdR = 0\,.
\end{align}
Because~$\epsilon>0$, and assuming~$f$ to be differentiable, we can
apply L'H\^opital's rule to the LHS of equation~\eqref{interror1} in
order to get,
\begin{align}\label{interror2}
	(-n+1-\epsilon)^{-1}\lim_{R\rightarrow \infty}R^{n+\epsilon}f\,.
\end{align}
By definition we have that~$f=o^+(R^{-n})=o(R^{-n-\epsilon})$, which
implies directly that~\eqref{interror2} is
zero. Therefore,~\eqref{interror3} must be true. All error terms in
this work are of the form~$o^+(R^{-n})$, therefore this result will be
used in every integration thereof. Note that if we had made the less
restrictive assumption that~$f=o(R^{-n+1})$ instead
of~$f=o^+(R^{-n+1})$, for the case where~$n=1$ we would have had to
show that,
\begin{align}\label{interror4}
	\lim_{R\rightarrow \infty}\int fdR = 0\,.
\end{align}
This would not be possible using the same method and we would not be
able to ensure that error terms remain small.

\paragraph*{Induction proof:} We have seen that,
\begin{align}\label{gn-1}
	g = \frac{G_1(\psi^*)}{R} + \frac{\mathcal{G}_2}{R^2}\,,
\end{align}
with~$\mathcal{G}_2=o^+(R)$, so in order to prove our
result~\eqref{gn}, we only need to show that if~$g$ can be written as,
\begin{align}\label{gn-1}
  g = \sum_{m=1}^{n-1} \frac{G_m(\psi^*)}{R^m}
  + \frac{\mathcal{G}_n}{R^{n}}\,,
\end{align}
with $\mathcal{G}_n=o^+(R)$, then it can be written as,
\begin{align}\label{gn2}
  g = \sum_{m=1}^{n} \frac{G_m(\psi^*)}{R^m}
  + \frac{\mathcal{G}_{n+1}}{R^{n+1}}\,,
\end{align}
with~$\mathcal{G}_{n+1}=o^+(R)$. We assume that each term
in~\eqref{gn-1} satisfies~\eqref{weaknull1}, so we have a way to
collect terms of the same order. Then we plug~\eqref{gn-1}
in~\eqref{gbu} and formally equate terms of order~$R^{-n-1}$ to find
the PDE,
\begin{align}\label{goodflatrecursion}
  2R^{n}\nabla_{\psi}\left(\frac{1}{R^{n-1}}\nabla_T\mathcal{G}_n\right)
  \simeq& \left[(n-1)(n-2)+\tilde{\cancel{\Delta}}\right]G_{n-1}\,.
\end{align}
Integrating equation~\eqref{goodflatrecursion} we find,
\begin{align}\label{goodflatrecursion4}
\mathcal{G}_n \simeq& -\frac{1}{2}
\left[n-2+\frac{\tilde{\cancel{\Delta}}}{n-1}\right]
\int_{T_0}^{T} G_{n-1}dT' \non\\
&+ R^{n-1}\int_{T_0}^{T} \dot{g}_n(\psi^*)dT'+ m_{g,n}\,.
\end{align}
It can be seen from~\eqref{goodflatrecursion4} that, to leading
order,~$\mathcal{G}_n$ does not satisfy~\eqref{gn-1} for a
general~$\dot{g}_n$. However,~$\dot{g}_n$ comes with a factor of
$R^{n-1}$ so, exactly like in~\eqref{g2=0}, the behavior of those
terms is captured by~$\dot{g}_1$. We can then choose solutions with
initial data such that~$\dot{g}_n=0\,,\forall n>1$ without loss of
generality. Putting aside for a moment the fact
that~\eqref{goodflatrecursion4} violates~\eqref{gn-1}, we could write
the~$T$ derivative of~$g$ to leading order as,
\begin{align}
  \nabla_Tg\simeq\frac{1}{R}(\dot{g}_1+...+\dot{g}_n)
  +\frac{1}{2}\tilde{\cancel{\Delta}}G_1\,.
\end{align}
Notice that all the~$\dot{g}_n$ appear with the same prefactor. So the
freedom that the integrations along $\psi$ at each order give us in
choosing initial data can in fact be expressed in the choice of one
scalar function. This is expected because the fact that we have~$n$
differential equations to solve for the different orders in the
field~$g$ is somewhat artificial, in the sense that they arise from a
method to solve a single differential equation. It is therefore
natural that once we add all the terms, we are left with only one free
function.  The same approach will be taken for the fields~$b$ and~$u$
whenever an integration along integral curves of~$\psi^a$ is
done. This gives the result,
\begin{align}\label{goodflatrecursion3}
  \mathcal{G}_n \simeq& -\frac{1}{2}\left[n-2
    +\frac{\tilde{\cancel{\Delta}}}{n-1}\right]
  \int_{T_0}^{T} G_{n-1}dT' + m_{g,n}\,.
\end{align}
Clearly,~\eqref{gn-1} implies~\eqref{gn2}, so we conclude that~$g$ can
be written as~\eqref{gn}. Moreover, up to $m_{g,n}$, we have a
recursion relation that allows us to compute~$G_n$ from~$G_{n-1}$ for
any~$n>1$,
\begin{align}\label{goodflatrecursionfinal}
  G_n=& -\frac{1}{2}\left[n-2+\frac{
      \tilde{\cancel{\Delta}}}{n-1}\right]
  \int_{T_0}^{T} G_{n-1}dT' + m_{g,n}\,.
\end{align}

\subsection{The Bad field}

\paragraph*{Motivation for induction hypothesis:} The case
of~$b$ requires a different hypothesis. Following the same kind of
procedure as in the~$g$ case, we begin by rescaling~$b$ in the
following way,
\begin{align}\label{bresc}
	\mathcal{B}_1:=bR\,,
\end{align}
and plugging it into~\eqref{toybox1}. The~$b$ equation in~\eqref{gbu}
can then be written as,
\begin{align}\label{bmot1}
  -2\nabla_\psi\nabla_T\mathcal{B}_1
  +\nabla_\psi^2\mathcal{B}_1
  +\cancel{\Delta}\mathcal{B}_1 = R(\nabla_Tg)^2\,,
\end{align}
where~$g$ is now given. The bad field satisfies~\eqref{weaknull1}, so
collecting terms of the lowest non-trivial order gives the following,
\begin{align}\label{bmot2}
  \nabla_\psi\nabla_T\mathcal{B}_1
  \simeq -\frac{1}{2R}(\nabla_TG_1)^2\,.
\end{align}
We can integrate~\eqref{bmot2} to get,
\begin{align}\label{bmot3}
  \mathcal{B}_1 \simeq -\frac{1}{2}\log R
  \int_{T_0}^{T}(\nabla_TG_1)^2 dT'
  + \int_{T_0}^{T} \dot{b}_1(\psi^*)dT'+m_{b,1}\,,
\end{align}
where~$\dot{b}_1(\psi^*)$ is a scalar function. The behavior of~$b$
to leading order differs from that of~$g$ as it has a term that grows
with $\log R$. This result is in accordance
with~\cite{GasGauHil19}. We define the function~$\mathcal{B}_2$ as,
\begin{align}\label{B'}
\frac{\mathcal{B}_2}{R} := \mathcal{B}_1 - B_1 =o^+(1)\,,
\end{align}
where,
\begin{align}\label{B1}
&B_1:=B_{1,0}(\psi^*)+ B_{1,1}(\psi^*)\log R\,,\non\\
&B_{1,0}(\psi^*) := \int_{T_0}^{T} \dot{b}_1(\psi^*)dT'+m_{b,1}\,,\non\\
&B_{1,1}(\psi^*) := -\frac{1}{2}\int_{T_0}^{T}(\nabla_TG_1)^2 dT'\,,
\end{align}
and assume that it also satisfies~\eqref{weaknull1}. The
subscripts~$n$ and~$k$ in~$B_{n,k}$ stand for the power of~$R^{-1}$
and the power of~$\log R$ associated with~$B_{n,k}$, respectively, and
the same notation will be used throughout this work except in the case
of~$g$ for a flat metric, where it is obvious that the associated
field vanishes when~$k$ is non-zero. While this notation seems
needlessly cumbersome at this point, it will prove useful in the next
section, where we find various combinations of powers of~$R^{-1}$
and~$\log R$. Replacing~\eqref{B'} in~\eqref{bmot1} and equating
lowest order terms gives,
\begin{align}
  2\nabla_\psi\left(\frac{1}{R}\nabla_T\mathcal{B}_2\right)
  + \frac{1}{R^2}B_{1,1} - \cancel{\Delta}B_1  \simeq
  -\frac{2}{R^2}\nabla_TG_1\nabla_TG_2\,,\non
\end{align}
which we can integrate to get,
\begin{align}\label{b'2}
\mathcal{B}_2& \simeq \frac{1}{2}
(1 - \tilde{\cancel{\Delta}})\int_{T_0}^T B_{1,1}dT'
- \frac{1}{2}\tilde{\cancel{\Delta}}\int_{T_0}^T
B_{1} dT'+m_{b,2} \non\\
&+ \int_{T_0}^T \nabla_TG_1\nabla_TG_2dT'\,,
\end{align}
meaning we can write that,
\begin{align}\label{B'3}
\mathcal{B}_2 \simeq B_{2,0}(\psi^*) + B_{2,1}(\psi^*)\log R \,.
\end{align}
Equations~\eqref{bmot3} and~\eqref{B'3} suggest that the field $b$ may
be written in the form,
\begin{align}\label{bn}
	b = \sum_{n=1}^\infty \frac{B_n}{R^n}\,,
\end{align}
with~$B_n=B_{n,0}(\psi^*)+B_{n,1}(\psi^*)\log R $ and we prove this
result in the following.

\paragraph*{Induction proof:} We have computed the bad field to
first order,
\begin{align}
	b = \frac{B_1}{R} + \frac{\mathcal{B}_2}{R^2}\,,
\end{align}
with~$\mathcal{B}_2=o^+(R)$, so in order to prove our
result~\eqref{gn}, we must show that if~$b$ can be written as,
\begin{align}\label{bn-1}
  b = \sum_{m=1}^{n-1} \frac{B_m}{R^m}
  + \frac{\mathcal{B}_n}{R^{n}}\,,
\end{align}
with~$\mathcal{B}_n=o^+(R)$, then it can be written as,
\begin{align}\label{bn2}
  b = \sum_{m=1}^{n} \frac{B_m}{R^m}
  + \frac{\mathcal{B}_{n+1}}{R^{n+1}}\,,
\end{align}
with~$\mathcal{B}_{n+1}=o^+(R)$. We assume all terms in the sum
in~\eqref{bn-1}, as well as~$\mathcal{B}_n$, satisfy~\eqref{weaknull1}
and we plug~\eqref{bn-1} into~\eqref{gbu} to get,
\begin{align}\label{badflatrecursion}
2R^{n}&\nabla_{\psi}\left(\frac{1}{R^{n-1}}
\nabla_T\mathcal{B}_n\right) \simeq -(2n-3)B_{n-1,1}\non\\
& + \left[(n-1)(n-2)+\tilde{\cancel{\Delta}}\right]B_{n-1} - C_n\,,
\end{align}
where~$C_n$ is defined as,
\begin{align}
	C_n := \sum_{i,j=1}^{i+j=n+1}\nabla_TG_i\nabla_TG_j\,.
\end{align}
Here,~$\sum_{i,j=1}^{i+j=n+1}$ is meant as the \textit{sum over terms
  with any combination of~$i$ and~$j$ as long as~$i,j\geq 1$
  and~$i+j=n+1$}. This can be integrated to get,
\begin{align}
  &\mathcal{B}_n \simeq \frac{1}{2}
  \left[1 - \frac{\tilde{\cancel{\Delta}}}{(n-1)^2}
    \right]\int_{T_0}^T B_{n-1,1}dT'+m_{b,n} \\
  &- \frac{1}{2}\left[n-2+\frac{\tilde{
        \cancel{\Delta}}}{n-1}\right]\int_{T_0}^T
  B_{n-1} dT'+ \frac{1}{2(n-1)}\int_{T_0}^T C_ndT'\,,\non
\end{align}
as we wanted. As with~$\dot{g}_n$ above, we choose the
functions~$\dot{b}_n=0\,,\forall n>1$, effectively absorbing them
into~$\dot{b}_1$ in order to avoid a contradiction with
assumptions~\eqref{bn-1}. Finally we get a recursion relation
for~$B_n$,
\begin{align}\label{badflatrecursionfinal}
  &B_n = \frac{1}{2}\left[1 - \frac{\tilde{\cancel{\Delta}}}
    {(n-1)^2}\right]\int_{T_0}^T B_{n-1,1}dT'+m_{b,n}\\
  &- \frac{1}{2}\left[n-2+\frac{\tilde{\cancel{\Delta}}}
    {n-1}\right]\int_{T_0}^T B_{n-1} dT'
  + \frac{1}{2(n-1)}\int_{T_0}^T C_ndT'\,.\non
\end{align}
This shows our hypothesis~\eqref{bn}, with,
\begin{align}
  &B_{n,0} = \frac{1}{2}\left[1
    - \frac{\tilde{\cancel{\Delta}}}{(n-1)^2}\right]
  \int_{T_0}^T B_{n-1,1}dT'+m_{b,n} \\
  &- \frac{1}{2}\left[n-2
    +\frac{\tilde{\cancel{\Delta}}}{n-1}\right]
  \int_{T_0}^T B_{n-1,0} dT'+ \frac{1}{2(n-1)}
  \int_{T_0}^T C_ndT'\,.\non
\end{align}
and,
\begin{align}
  B_{n,1} =- \frac{1}{2}\left[n-2
    +\frac{\tilde{\cancel{\Delta}}}{n-1}\right]\int_{T_0}^T B_{n-1,1} dT'\,,
\end{align}
A closer look at~\eqref{badflatrecursionfinal} reveals that the
only~$\log R$ term comes from~$B_{n-1}$, and hence it is inherited by
all orders from~$B_1$.

\subsection{The Ugly field}

\paragraph*{Motivation for induction hypothesis:} Once again we
rescale the field $u$ by $R$ as,
\begin{align}
	\mathcal{U}_1=uR\,,
\end{align}
and plug it into~\eqref{toybox1}. The $u$ equation in~\eqref{gbu} can
then be written as,
\begin{align}\label{umot1}
  -\frac{2}{R}\nabla_\psi\left(R\nabla_T\mathcal{U}_1\right)
  + \nabla_\psi^2\mathcal{U}_1 +\cancel{\Delta}\mathcal{U}_1 = 0\,,
\end{align}
The ugly field has a somewhat different behavior from the other
two. As can be seen in~\cite{GasGauHil19}, both good and bad
derivatives improve the fall-off of $u$, as opposed to the cases
of~$g$ and~$b$, where only good derivatives improve. This means that
all terms in~\eqref{umot1} contribute to leading order and one would
have to solve the whole equation at once. For this reason, we will
focus on solutions that can be decomposed as,
\begin{align}\label{umot4}
	\mathcal{U}_1 = m_{u,1} + \frac{\mathcal{U}_2}{R}\,,
\end{align}
where~$m_{u,1}$ is independent of $T$ and $R$ and $\mathcal{U}_2 =
o^+(R)$. Plugging this into~\eqref{umot1} we get,
\begin{align}\label{umot2}
  -\frac{2}{R}\nabla_\psi\nabla_T\mathcal{U}_2
  + \left(\nabla_\psi^2 + \cancel{\Delta}\right)
  \left(m_{u,1}+\frac{\mathcal{U}_2}{R}\right) = 0\,.
\end{align}
If we assume~$\mathcal{U}_2$ to satisfy~\eqref{weaknull1},
\begin{align}\label{umot3}
  &2\nabla_\psi\nabla_T\mathcal{U}_2 \simeq
  \frac{1}{R}\tilde{\cancel{\Delta}}m_{u,1}\Rightarrow \\
  &\mathcal{U}_2 \simeq \int_{T_0}^{T}\left[\log
    R \tilde{\cancel{\Delta}}m_{u,1}
    + \dot{u}_2(\psi^*)\right]dT' + m_{u,2}\,.\non
\end{align}
Equation~\eqref{umot3} suggests that the ugly field can be written as,
\begin{align}\label{un}
	u = \frac{m_{u,1}}{R} + \sum_{n=2}^\infty \frac{U_n}{R^n}\,,
\end{align}
with~$U_n= U_{n,0}(\psi^*)+U_{n,1}(\psi^*)\log R$ and we show this
result in the following.

\paragraph*{Induction proof:} As was seen above, the first order
term of the ugly field behaves differently from the rest, in that {\it
  all} of its derivatives improve. For this reason we begin the
induction proof in the second order term, which has been computed
in~\eqref{umot4} and~\eqref{umot3}. To prove~\eqref{un} we have to
show that if,
\begin{align}\label{un-1}
  u = \frac{m_{u,1}}{R}+\sum_{m=2}^{n-1} \frac{U_m}{R^m}
  + \frac{\mathcal{U}_n}{R^{n}}\,,
\end{align}
with~$\mathcal{U}_n=o^+(R)$, then,
\begin{align}\label{un2}
  u = \frac{m_{u,1}}{R}+\sum_{m=2}^{n} \frac{U_m}{R^m}
  + \frac{\mathcal{U}_{n+1}}{R^{n+1}}\,,
\end{align}
with~$\mathcal{U}_{n+1}=o^+(R)$. We assume that all orders
in~\eqref{un-1} and~$\mathcal{U}_n$ satisfy~\eqref{weaknull1} and we
substitute that in~\eqref{gbu} to get,
\begin{align}\label{uglyflatrecursion2}
  2R^{n-1}\nabla_{\psi}\left(\frac{1}{R^{n-2}}
  \nabla_T\mathcal{U}_n\right) = -(2n-3)U_{n-1,1} \non\\
	+ \left[(n-1)(n-2)+\tilde{\cancel{\Delta}}\right]U_{n-1}\,,
\end{align}
which we can integrate to get, $\forall n>2$,
\begin{align}\label{uglyflatrecursion3}
  \mathcal{U}_n \simeq& \frac{1}{2}
  \left[1 - \frac{\tilde{\cancel{\Delta}}}{(n-1)^2}
    \right]\int_{T_0}^T U_{n-1,1}dT'+m_{u,n} \\
  &- \frac{1}{2}\left[n-2+\frac{\tilde{\cancel{\Delta}}}
    {n-1}\right]\int_{T_0}^T U_{n-1} dT' \non\,.
\end{align}
As in the~$g$ and~$b$ cases, we consider the initial data arising from
the~$\psi^a$ integration~$\dot{u}_n$ to be zero for all~$n>2$. This
concludes the proof by induction and we get a final recursion relation
for~$U_n$ in terms of~$U_{n-1}$,
\begin{align}\label{uglyflatrecursionfinal}
  U_n =& \frac{1}{2}\left[1 - \frac{\tilde{\cancel{\Delta}}}
    {(n-1)^2}\right]\int_{T_0}^T U_{n-1,1}dT'+m_{u,n} \\
  &- \frac{1}{2}\left[n-2+\frac{\tilde{\cancel{\Delta}}}
    {n-1}\right]\int_{T_0}^T U_{n-1} dT' \non\,.
\end{align}
Our results are summarized by the following:

\begin{thm}
  Let~$X^{\ul{\alpha}}=(T,X^{\ul{i}})$ be an asymptotically Cartesian
  coordinate system with an associated covariant derivative~$\mn$. The
  \textit{good-bad-ugly} system defined as,
  \begin{align}
    &\mathring{\square} g = 0\,, \non\\
    &\mathring{\square} b = (\nabla_T g)^2\,,\non\\
    &\mathring{\square} u = \tfrac{2}{R}\nabla_T u\,,
  \end{align}
  where~$\mathring{\square}:=\eta^{ab}\mn_a\mn_b$ and~$\eta$ is the
  Minkowski metric, admits formal polyhomogeneous asymptotic solutions
  near null infinity of the type,
  \begin{align}
    &g= \sum_{n=1}^{\infty}\frac{G_n(\psi^*)}{R^n}\non\,,\\
    &b= \sum_{n=1}^\infty \frac{B_n}{R^n}\,,\non\\
    &u= \frac{m_{u,1}}{R} + \sum_{n=2}^\infty \frac{U_n}{R^n}\,,
  \end{align}
  where~$B_n=B_{n,0}(\psi^*)+B_{n,1}(\psi^*)\log R$
  and~$U_n=U_{n,0}(\psi^*)+U_{n,1}(\psi^*)\log R$ and with initial
  data on~${\mathcal{S}}$ of the type,
  \begin{align}
    \begin{cases}
      g\rvert_{\mathcal{S}}= \sum_{n=1}^{\infty}\frac{m_{g,n}}{R^n}\\
      b\rvert_{\mathcal{S}}= \sum_{n=1}^\infty \frac{m_{b,n}}{R^n}\\
      u\rvert_{\mathcal{S}}= \sum_{n=1}^\infty \frac{m_{u,n}}{R^n}
    \end{cases}\,,
    \begin{cases}
      \nabla_Tg\rvert_{\mathcal{S}}= O_{\mathcal{S}}(R^{-2})\\
      \nabla_Tb\rvert_{\mathcal{S}}= O_{\mathcal{S}}(R^{-2})\\
      \nabla_Tu\rvert_{\mathcal{S}}= O_{\mathcal{S}}(R^{-2})
    \end{cases}\,,
	\end{align}
  where~$m_{\phi,n}$ are scalar functions that are independent of~$T$
  and~$R$. This is valid outside a compact ball centered
  at~$R=0$. Additionally, the functions~$G_n$ are given by~\eqref{G1}
  and~\eqref{goodflatrecursionfinal}, $B_n$ are given by~\eqref{B1}
  and~\eqref{badflatrecursionfinal} and~$U_n$ by~\eqref{umot3}
  and~\eqref{uglyflatrecursionfinal}.
\end{thm}

\begin{remark} Looking at equation~\eqref{uglyflatrecursionfinal}
  we see that the only way for~$U_n$ to have a term with~$\log R$ in
  it is if~$U_{n-1}$ does too. As this is valid for any~$n>2$, the
  orders of~$U_n$ higher than~$U_2$ can only have a~$\log R$ term if
  $U_2$ does as well. In other words, if we
  require~$\cancel{\Delta}m_{u,1}=0$, the field~$u$ will have no~$\log
  R$ terms at any order. In fact, with that requirement, it can easily
  be seen that~$U_n$ satisfies a hypothesis analogous to that of the
  field~$g$, namely,
\begin{align}
	u=\sum_{n=1}^{\infty}\frac{U_n(\psi^*)}{R^n}\,.
\end{align}
\end{remark}

\begin{remark} The \textit{good-bad-ugly} system~\eqref{gbu} admits
  a static solution that is obtained with the following initial data
  in~$\mathcal{S}$,
  \begin{align}\label{static1}
    \begin{cases}
      g\rvert_{\mathcal{S}}= \frac{m_{g,1}}{R}\\
      b\rvert_{\mathcal{S}}= \frac{m_{b,1}}{R}\\
      u\rvert_{\mathcal{S}}= \frac{m_{u,1}}{R}
    \end{cases}
  \end{align}
where the functions~$m_{\phi,1}$ satisfy the
condition~$\cancel{\Delta}m_{\phi,1}=~0$. In that case, looking at the
recursion relations~\eqref{goodflatrecursionfinal},
\eqref{badflatrecursionfinal} and~\eqref{umot3}, we see that the
series is truncated at~$n=1$ and hence all the higher order terms
vanish.
\end{remark}

\section{Asymptotically flat metrics}\label{section:gencase}

In this section we follow the same procedure as in the prequel, but
this time employing a more general metric whose functions are allowed
to depend analytically on the fields~$g, b$ and~$u$. Since we want to
maintain the generality of those functions, the final recursion
relation for the \textit{good-bad-ugly} system will have to be written
as a function of~$g$, $b$ and~$u$. Nevertheless, we will see that an
induction proof analogous to that of the flat metric case can be made
for a general asymptotically flat metric under the assumptions given in
section~\ref{assumptions}. An asymptotically flat metric that is
simply given, rather than occurring as a function of our unknown
fields, could be treated similarly.

\paragraph*{Expansion of the reduced wave operator:} Let~$\phi$ be
any field in~$\{g,b,u\}$ and use~\eqref{eq:metricrepresentation} to
expand the LHS of~\eqref{gbu},
\begin{align}\label{lhs1}
    \mathring{\square} \phi =
    &\left[-\frac{2e^{-\varphi}}{\tau}\nabla_{\psi}\nabla_{\ul{\psi}}
      + \frac{2e^{-\varphi}}{\tau}(\mn_{\psi} \ul{\psi})^a\nabla_a-
      \frac{1}{\tau}\mathring{\sD}^a\sigma_a\nabla_{\ul{\psi}}\right.\non
      \\ &\left. -
      \frac{1}{\tau}\mathring{\sD}^a\ul{\sigma}_a\nabla_{\psi} +
      \cancel{\Delta}\right]\phi\,.
\end{align}
We want to write expression~\eqref{lhs1} in terms of derivatives of
metric functions and~$\phi$ along the vector fields~$\psi^a$
and~$\p_T^a$. For clarity, let us treat each term individually and put
everything together in the end. Using~\eqref{eq:psiinshellchart}, the
first term on the RHS turns into,
\begin{align}
  \nabla_{\psi}\nabla_{\ul{\psi}}\phi =
  \nabla_\psi\left(\frac{\tau}{\mathcal{C}_+^R}\nabla_T\phi
  + \frac{\mathcal{C}_-^R}{\mathcal{C}_+^R}\nabla_\psi\phi\right) \,.
\end{align}
From the second term on the RHS we get,
\begin{align}
  (\mn_{\psi} \ul{\psi})^a\nabla_a\phi &=
  (\mbn_{\psi} \ul{\psi})^a\nabla_a\phi=
  \frac{1}{\mathcal{C}_+^R}\nabla_\psi
  \mathcal{C}_-^R(\nabla_\psi\phi - \nabla_T\phi)\,,\non
\end{align}
where the first equality comes from the fact
that~$\Gmmb_{\psi}{}^a{}_{\ul{\psi}}=0$. The third term can be
expanded as,
\begin{align}
  &\mathring{\sD}^a\sigma_a\nabla_{\ul{\psi}}\phi
  = \slashed{g}^{ab}(\mbn_a\sigma_b - \Gmmb_a{}^\sigma{}_{b})
  \nabla_{\ul{\psi}}\phi\\
  &= \slashed{g}^{ab}(-\nabla_bT\nabla_a\mathcal{C}^R
  + \nabla_b\theta^A\nabla_a\mathcal{C}^+_A - \Gmmb_a{}^\sigma{}_{b})
  \nabla_{\ul{\psi}}\phi\non\\
  &=\left(-\frac{\mathcal{C}_A}{\tau}\slashed{D}^A\mathcal{C}^R
  +\slashed{D}^A\mathcal{C}_A^+- \slashed{g}^{ab}\Gmmb_{a}{}^{\sigma}{}_{b}\right)
  \nabla_{\ul{\psi}}\phi\non\,,
\end{align}
where~$\mathcal{C}_A:=\mathcal{C}^+_A+\mathcal{C}^-_A$ and
$\nabla_{\ul{\psi}}\phi$ should be written in terms of
$\nabla_{T}\phi$ and $\nabla_{\psi}\phi$
with~\eqref{eq:psiinshellchart}, whereas the fourth term reads,
\begin{align}
  \mathring{\sD}^a&\ul{\sigma}_a\nabla_{\psi}\phi
  = \left(-\frac{\mathcal{C}_A}{\tau}\slashed{D}^A
  \mathcal{C}^R+\slashed{D}^A\mathcal{C}_A^--
  \slashed{g}^{ab}\Gmmb_{a}{}^{\ul{\sigma}}{}_{b}\right) \nabla_{\psi}\phi\non\,.
\end{align}
Putting all of this together in~\eqref{lhs1} gives,
\begin{align}\label{lhs2}
  \mathcal{C}_+^R\mathring{\square} \phi = &-
  2e^{-\varphi}\nabla_{\psi}\nabla_{T} \phi
  +\nabla_T\phi(\slashed{g}^{ab}\Gmmb_{a}{}^{\sigma}{}_{b} + X_T)\non\\
  &+\nabla_{\psi}\phi X_\psi
  -\frac{e^{-\varphi}\mathcal{C}_-^R}{\tau}\nabla_{\psi}^2\phi
  + \mathcal{C}_+^R\cancel{\Delta}\phi\,,
\end{align}
where~$X_T$ and~$X_\psi$ are,
\begin{align}
  \tau X_T:= &\mathcal{C}_A\sD^A\mathcal{C}_-^R
  - \tau\sD^A\mathcal{C}_A^+
  + \frac{2e^{-\varphi}\mathcal{C}_-^R}{\mathcal{C}_+^R}
  \nabla_\psi\mathcal{C}_+^R\non\,,\\
  \tau X_\psi:=& \frac{\mathcal{C}_A}{\tau}
  \sD^A(\mathcal{C}_-^R\mathcal{C}_+^R)
  - \mathcal{C}_-^R\sD^A\mathcal{C}_A^+
  - \mathcal{C}_+^R\sD^A\mathcal{C}_A^- \\\non
  + \mathcal{C}_-^R&\slashed{g}^{ab}\Gmmb_{a}{}^{\sigma}{}_{b}
  + \mathcal{C}_+^R\slashed{g}^{ab}\Gmmb_{a}{}^{\ul{\sigma}}{}_{b}
  + \frac{2e^{-\varphi}\mathcal{C}_-^R}{\mathcal{C}_+^R}
  \nabla_\psi\mathcal{C}_+^R \,.
\end{align}

\paragraph*{Motivation for induction hypothesis:} As the~$\gamma$
functions (see \eqref{asympflatness}) are analytic functions of the
evolved fields at null infinity, we can Taylor expand them
around~$g=b=u=0$, because the fields are assumed to have decay near
null infinity. That gives,
\begin{align}\label{taylor}
  \gamma(g,b,u)&= \sum_{i=0}^\infty\sum_{j=0}^\infty\sum_{k=0}^\infty
  \frac{g^i b^j u^k}{i!j!k!}\left(\frac{\p^{i+j+k}\gamma}
       {\p g^i\p b^j\p u^k}\right)\Big|_{\mathscr{I}^+}\non\\
       &=\frac{\p \gamma}{\p g}\Big|_{\mathscr{I}^+}g
       +\frac{\p \gamma}{\p b}\Big|_{\mathscr{I}^+}b
       +\frac{\p \gamma}{\p u}\Big|_{\mathscr{I}^+}u+...\,,
\end{align}
where the second equality uses the fact
that~$\gamma|_{\mathscr{I}^+}=0$, because the metric is asymptotically
flat. Equation~\eqref{taylor} then implies that~$\gamma=o^+(1)$ and,
\begin{align}\label{dersgamma}
  \omega_\gamma = o^+(R^{-n})\Rightarrow
  \begin{cases}
    \nabla_{\psi}\omega_\gamma = o^+(R^{-n-1})\\
    \nabla_{X_A}\omega_\gamma = o^+(R^{-n-1})
  \end{cases}
  \,,
\end{align}
where~$\omega_\gamma$ is any~$\gamma$ function or any derivative of
it. Note that we intentionally left out any bad derivatives because in
order to know the asymptotic behavior of those we would have to
specify the dependence of~$\gamma$ on~$g$, $b$ and~$u$. Remarkably,
one can easily check that the expanded form of the reduced wave
operator~\eqref{lhs2} does not include any bad derivatives of metric
functions. Let us rescale~$g$, $b$ and~$u$
as~\eqref{gresc},~\eqref{bresc} and~\eqref{umot4},
respectively. With~\eqref{dersgamma} one can count the order of each
term in~\eqref{lhs2} and see that only the first two terms contribute
to leading order. These are exactly the same terms that contribute to
leading order in the flat metric case, which means that none of the
extra terms that arise from allowing the spacetime to have curvature
can possibly contribute to first order. We get the equations,
\begin{align}
  &\nabla_\psi\nabla_T\mathcal{G}_1 \simeq 0\,,\non\\
  &\nabla_\psi\nabla_T\mathcal{B}_1 \simeq
  -\frac{2}{R}(\nabla_TG)^2\,,\non\\
  &2\nabla_\psi\nabla_T\mathcal{U}_2 \simeq
  \frac{1}{R}\tilde{\cancel{\Delta}}m_{u,1}\,.
\end{align}
Note that, to leading order,~$\tilde{\cancel{\Delta}}$ is the
Laplacian on the $2$-sphere of unit radius,
\begin{align}
  \tilde{\cancel{\Delta}}\phi \simeq R^2
  {\slashed{\eta}}^{ab}\mn_a
  \left({\slashed{\eta}}^c{}_b\mn_c\phi\right)\,,
\end{align}
because~$\slashed{g}^{ab}$ approaches the inverse
metric on the 2-sphere of radius~$R$. Therefore we have,
\begin{align}\label{g1b1u1}
&\mathcal{G}_1 \simeq G_{1,0}(\psi^*)\non\\
&\mathcal{B}_1 \simeq B_{1,0}(\psi^*) + B_{1,1}(\psi^*)\log R\non\\
&\mathcal{U}_2 \simeq \int_{T_0}^{T}
\left[\log R \tilde{\cancel{\Delta}}m_{u,1}
  + \dot{u}_2(\psi^*)\right]dT' + m_{u,2}(T_0)\,.
\end{align}
In~\eqref{g1b1u1} we kept~$\mathcal{U}_2$ as a function of~$m_{u,1}$
because there is a remark to be made about it at the end of this
section. As the metric functions are free to depend upon the evolved
fields, the second order equations (third order in the case of~$u$)
may be coupled to first order terms. This means
that~$\mathcal{G}_1-G_{1,0}$, for instance, could have a term
proportional to~$\log R$ that is coming from~$\mathcal{B}_1$. On the
other hand, non-linearities could give rise to terms proportional to
higher powers of~$\log R$. This seems to suggest that~$g$, $b$ and~$u$
are polyhomogeneous functions where each term can have up to $n$
powers of~$\log R$ in the~$b$ case, and up to~$n-1$ in the~$g$ and~$u$
cases. Formally, we therefore conjecture
\begin{align}\label{hyp}
  &g = \sum_{n=1}^{\infty}\sum_{k=0}^{n-1} \frac{(\log R)^k
    G_{n,k}(\psi^*)}{R^n}\non\\
  &b =  \sum_{n=1}^{\infty}\sum_{k=0}^{n} \frac{(\log R)^k
    B_{n,k}(\psi^*)}{R^n}\\
  &u =  \frac{m_{u,1}}{R}+\sum_{n=2}^{\infty}\sum_{k=0}^{n-1}
  \frac{(\log R)^k U_{n,k}(\psi^*)}{R^n}	\non\,.
\end{align}
We proceed by induction as in the the previous
cases. From~\eqref{g1b1u1} we can already know that to first order
in~$g$ and~$u$,~$\log R$ terms are not allowed, and the
conjecture~\eqref{hyp} incorporates this property by
construction. Truncating at~$n=1$, we have seen
\begin{align}
  &g = \frac{G_{1,0}(\psi^*)}{R}
  + \frac{\mathcal{G}_2}{R^2}\,,\non\\
  &b = \frac{B_{1,0}(\psi^*) + B_{1,1}(\psi^*)\log R}{R}
  + \frac{\mathcal{B}_2}{R^2}\,,\\
  &u = \frac{m_{u,1}}{R}+\frac{U_{2,0}(\psi^*)
    + U_{2,1}(\psi^*)\log R}{R^2} +\frac{\mathcal{U}_3}{R^3} \non\,,
\end{align}
with~$\mathcal{G}_2=o^+(R)$, $\mathcal{B}_2=o^+(R)$
and~$\mathcal{U}_3=o^+(R)$, so in order to show~\eqref{hyp}, we have
to show that if we can write the evolved fields as,
\begin{align}\label{phin-1}
  &g = \sum_{m=1}^{n-1}\sum_{k=0}^{m-1} \frac{(\log R)^k
    G_{m,k}(\psi^*)}{R^m}+\frac{\mathcal{G}_n}{R^n}\non\\
  &b =  \sum_{m=1}^{n-1}\sum_{k=0}^{m} \frac{(\log R)^k
    B_{m,k}(\psi^*)}{R^m}+\frac{\mathcal{B}_n}{R^n}\\
  &u =  \frac{m_{u,1}}{R}+\sum_{m=2}^{n-1}\sum_{k=0}^{m-1}
  \frac{(\log R)^k U_{m,k}(\psi^*)}{R^m}
  +\frac{\mathcal{U}_n}{R^n}\non\,,
\end{align}
where~$\mathcal{G}_n=o^+(R)$, $\mathcal{B}_n=o^+(R)$
and~$\mathcal{U}_n=o^+(R)$, then we can also write them as,
\begin{align}\label{phin}
  &g = \sum_{m=1}^{n}\sum_{k=0}^{m-1} \frac{(\log R)^k
    G_{m,k}(\psi^*)}{R^m}+\frac{\mathcal{G}_{n+1}}{R^{n+1}}\non\\
  &b =  \sum_{m=1}^{n}\sum_{k=0}^{m} \frac{(\log R)^k
    B_{m,k}(\psi^*)}{R^m}+\frac{\mathcal{B}_{n+1}}{R^{n+1}}\\
  &u =  \frac{m_{u,1}}{R}+\sum_{m=2}^{n}\sum_{k=0}^{m-1}
  \frac{(\log R)^k U_{m,k}(\psi^*)}{R^m}
  +\frac{\mathcal{U}_{n+1}}{R^{n+1}}	\,,\non
\end{align}
where~$\mathcal{G}_{n+1}=o^+(R)$, $\mathcal{B}_{n+1}=o^+(R)$
and~$\mathcal{U}_{n+1}=o^+(R)$. To do this, we must first find what
the metric functions, and hence the~$\gamma$ functions, behave like if
we assume~\eqref{phin-1}.

\paragraph*{Behavior of $\gamma$ functions:} Let~$\phi_1$
and~$\phi_2$ be any of the fields in~$\{g,b,u\}$. According to our
assumption~\eqref{phin-1}, $\phi_i$ with~$i\in\{1,2\}$ can be written
as
\begin{align}\label{phin-1.2}
		\phi_i = \sum_{m=1}^{n-1}\sum_{k=0}^m \frac{(\log R)^k
                  \Phi_{m,k}(\psi^*)}{R^m}+\frac{\Phi}{R^n}\,,
\end{align}
for suitable scalar functions~$\Phi_{m,k}(\psi^*)$
and~$\Phi=o^+(R)$. It is then straightforward to check that the
product of any two evolved fields can also be written as,
\begin{align}\label{phi2}
  \phi_1\phi_2 = \sum_{m=1}^{n-1}\sum_{k=0}^m \frac{(\log R)^k
    \bar{\Phi}_{m,k}(\psi^*)}{R^m}+\frac{\bar{\Phi}}{R^n}\,,
\end{align}
once again for suitable functions~$\bar{\Phi}_{m,k}(\psi^*)$
and~$\bar{\Phi}=o^+(R)$, which is formally the same
as~\eqref{phin-1.2}. This means that no matter how many times we
multiply any powers of the evolved fields, it is always possible to
write the resulting product as~\eqref{phi2}. If we plug~\eqref{phin-1}
into~\eqref{taylor} we get in each term a product of powers of the
fields~$g$, $b$ and~$u$, so we can write any~$\gamma$ function as,
\begin{align}\label{gamman-1}
  \gamma = \sum_{m=1}^{n-1}\sum_{k=0}^m
  \frac{(\log R)^k \Gamma_{m,k}(\psi^*)}{R^m}+\frac{\Gamma}{R^n}\,,
\end{align}
with~$\Gamma=o^+(R)$.

\paragraph*{Induction proof:} We plug~\eqref{phin-1}
and~\eqref{gamman-1} into~\eqref{lhs2} and collect terms proportional
to~$R^{-n-1}$. In the~$g$ equation, for instance, we see that the only
terms in~\eqref{lhs2} that may contain~$\mathcal{G}_n$ are the first
two, while none of the remaining terms may contain~$\mathcal{B}_n$
or~$\mathcal{U}_n$. Putting all terms
with~$\mathcal{G}_n$,~$\mathcal{B}_n$ and~$\mathcal{U}_n$ on the LHS
and all the rest on the RHS we get the system,
\begin{align}\label{recur1}
  &R^{n}\nabla_{\psi}
  \left(\frac{1}{R^{n-1}}\nabla_T\mathcal{G}_n\right)
  \simeq \Omega^g_{n-1}\,,\non \\
  &R^{n}\nabla_{\psi}\left(\frac{1}{R^{n-1}}\nabla_T\mathcal{B}_n\right)
  + \nabla_TG_n\nabla_TG_1 \simeq \Omega^b_{n-1}\,,\non\\
  &R^{n-1}\nabla_{\psi}\left(\frac{1}{R^{n-2}}
  \nabla_T\mathcal{U}_n\right) \simeq \Omega^u_{n-1}\,,
\end{align}
where on the right hand sides~$\Omega^{\phi}_{n-1}$ depend on the
functions~$\{G_{m,k},B_{m,k},U_{m,k},m_{u,1}\}$, for~$m\in[1,n-1]$
and~$k\in[0,m]$, and their
derivatives. Also,~$\Omega^{\phi}_{0}:=0$. At this point we need to
establish the maximum power of~$\log R$ in the
functions~$\Omega^{\phi}_{n-1}$. Since we are collecting terms of
order~$R^{-n-1}$, naively we would say that a term collected this way
could have a maximum power of~$n+1$. Although, any good derivative or
factor of~$R^{-1}$ increases the order of the term without increasing
the power of~$\log R$. For instance, in the third term of
equation~\eqref{lhs2} applied to~$b$, the maximum power is~$n$,
because it has one good derivative and no factors of~$R^{-1}$. With
this in mind, we can split the functions~$\Omega^{\phi}_{n-1}$ in the
following way,
\begin{align}\label{expH}
&\Omega^{g}_{n-1} = \sum_{p=0}^{n-1}(\log R)^p\Omega^{g}_{n-1,p}(\psi^*)\,,\non\\
&\Omega^{b}_{n-1} = \sum_{p=0}^{n}(\log R)^p\Omega^{b}_{n-1,p}(\psi^*)\,,\non\\
&\Omega^{u}_{n-1} = \sum_{p=0}^{n-1}(\log R)^p\Omega^{u}_{n-1,p}(\psi^*)\,.
\end{align}
It is worth noting that the specific form of
the~$\Omega^{\phi}_{n-1,p}$ functions has no influence on the proof of
our hypothesis, as long as it is possible to write~\eqref{expH}. In
fact~\eqref{expH} holds for a more general class of models than
just~\eqref{gbu}, as will be discussed in the next
section. Equation~\eqref{expH} allows us to integrate~\eqref{recur1}
in order to get the asymptotic behavior of~$\mathcal{G}_n$,
$\mathcal{B}_n$ and~$\mathcal{U}_n$ in terms
of~$\{G_{m,k},B_{m,k},U_{m,k},m_{u,1}\}$. Let us begin with the first
equation,
\begin{align}
  R^n\nabla_{\psi}\left(\frac{1}{R^{n-1}}
  \nabla_T\mathcal{G}_n\right) \simeq
  \sum_{p=0}^{n-1}(\log R)^p\Omega^g_{n-1,p}(\psi^*)\,.\non
\end{align}
We make use of the following integral,~$\forall q\neq 1$,
\begin{align}
  \int \frac{(\log R)^p}{R^q} dR &= -\frac{(\log R)^p}{(q-1)R^{q-1}}
  + \frac{p}{q-1}\int \frac{(\log R)^{p-1}}{R^q} dR\non\\
  &= \sum_{i=0}^p -\frac{(\log R)^i}{(q-1)^{p-i+1}R^{q-1}}\frac{p!}{i!}\,,
\end{align}
to get,
\begin{align}\label{ggeneral}
  \mathcal{G}_n \simeq& \sum_{p=0}^{n-1}\sum_{i=0}^p
  -\frac{(\log R)^i}{(n-1)^{p-i+1}}\frac{p!}{i!}\int_{T_0}^T
  \Omega^g_{n-1,p}dT'+ m_{g,n}(T_0)\non\\
  =& \sum_{i=0}^{n-1}(\log R)^i \sum_{p=0}^n
  -\frac{1}{(n-1)^{p-i+1}}\frac{p!}{i!}\int_{T_0}^T \Omega^g_{n-1,p}dT'\non\\
  &+ m_{g,n}(T_0)\non\\
  =& \sum_{i=0}^{n-1} (\log R)^i G_{n,i}(\psi^*)\,,
\end{align}
for some scalar functions~$G_{n,i}(\psi^*)$ and for all~$n> 1$. As in
the flat case, we choose~$\dot{g}_n=0$ for~$n>1$, and the same applies
for~$b$ (for $n>1$) and~$u$ (for~$n>2$). In other words, if~$g$ can be
written as~\eqref{phin-1}, then it can also be written
as~\eqref{phin}. Therefore, we have,
\begin{align}
  g = \sum_{n=1}^{\infty}\sum_{k=0}^{n-1}
  \frac{(\log R)^k G_{n,k}(\psi^*)}{R^n}\,,
\end{align}
as desired. The~$b$ equation likewise gives,
\begin{align}
  R^{n}\nabla_{\psi}\left(\frac{1}{R^{n-1}}
  \nabla_T\mathcal{B}_n\right) \simeq &
  - \nabla_TG_n\nabla_TG_{1,0} \\
  &+\sum_{p=0}^{n}(\log R)^p\Omega^b_{n-1,p}(\psi^*)\non\,,
\end{align}
which we can integrate in order to get,
\begin{align}\label{bgeneral}
  \mathcal{B}_n \simeq& \sum_{i=0}^n(\log R)^i \sum_{p=0}^n
  -\frac{1}{(n-1)^{p-i+1}}\frac{p!}{i!}\int_{T_0}^T
  \Omega^{b}_{n-1,p}dT'+ \non\\
  +&\sum_{i=0}^{n-1}(\log R)^i \sum_{p=0}^{n-1}
  -\frac{1}{(n-1)^{p-i+1}}\frac{p!}{i!}\non\\
  &\int_{T_0}^T \sum_{k=0}^{n-1} \nabla_TG_{n,k}
  \nabla_TG_{1,0}dT'+ m_{b,n}(T_0)\non\\
  =& \sum_{i=0}^{n} (\log R)^i B_{n,i}(\psi^*)\,,
\end{align}
for all~$n>1$. Thus, by induction, we get,
\begin{align}
  b = \sum_{n=1}^{\infty}\sum_{k=0}^{n}
  \frac{(\log R)^k B_{n,k}(\psi^*)}{R^n}\,.
\end{align}
Finally, the~$u$ equation reads,
\begin{align}
  R^{n-1}\nabla_{\psi}\left(\frac{1}{R^{n-2}}
  \nabla_T\mathcal{U}_n\right) \simeq
  \sum_{p=0}^{n-1}(\log R)^p\Omega^{u}_{n-1,p}(\psi^*)\,,\non
\end{align}
and integrating it along integral curves of~$\psi^a$ and~$\p_T^a$ gives,
\begin{align}\label{ugeneral}
  \mathcal{U}_n \simeq& \sum_{i=0}^{n-1}(\log R)^i \sum_{p=0}^{n-1}
  -\frac{1}{(n-2)^{p-i+1}}\frac{p!}{i!}\int_{T_0}^T
  \Omega^{U}_{n-1,p}dT'\non\\
  &+ m_{u,n}(T_0)\non\\
  =& \sum_{i=0}^{n-1} (\log R)^i U_{n,i}(\psi^*)\,,
\end{align}
for all~$n>2$. By induction,
\begin{align}
  u = \frac{m_{u,1}}{R}+\sum_{n=2}^{\infty}\sum_{k=0}^{n-1}
  \frac{(\log R)^k U_{n,k}(\psi^*)}{R^n}\,.
\end{align}
This concludes the proof. These results can be packaged in the
following theorem.

\begin{thm}\label{T2}
  Let $X^{\ul{\alpha}}=(T,X^{\ul{i}})$ be an asymptotically
  Cartesian coordinate system with an associated covariant
  derivative $\mn$. The \textit{good-bad-ugly} system defined
  as,
  \begin{align}
    \begin{cases}
      \mathring{\square} g = 0 \\
      \mathring{\square} b = (\nabla_T g)^2\\
      \mathring{\square} u = \frac{2}{R}\nabla_T u
    \end{cases}\,,
  \end{align}
  where $\mathring{\square}:=g^{ab}\mn_a\mn_b$ and $g$ is an
  asymptotically flat metric, admits a polyhomogeneous expansion
  near null infinity of the type,
  \begin{align}
    &g = \sum_{n=1}^{\infty}\sum_{k=0}^{n-1} \frac{(\log R)^k
      G_{n,k}(\psi^*)}{R^n}\non\\
    &b =  \sum_{n=1}^{\infty}\sum_{k=0}^{n} \frac{(\log R)^k
      B_{n,k}(\psi^*)}{R^n}\\
    &u =  \frac{m_{u,1}}{R}+\sum_{n=2}^{\infty}\sum_{k=0}^{n-1}
    \frac{(\log R)^k U_{n,k}(\psi^*)}{R^n}\non\,,
  \end{align}
  with initial data on~$\mathcal{S}$ of the type,
  \begin{align}
    \begin{cases}
      g\rvert_{\mathcal{S}}= \sum_{n=1}^{\infty}\frac{m_{g,n}}{R^n}\\
      b\rvert_{\mathcal{S}}= \sum_{n=1}^\infty \frac{m_{b,n}}{R^n}\\
      u\rvert_{\mathcal{S}}= \sum_{n=1}^\infty \frac{m_{u,n}}{R^n}
    \end{cases}\,,
  \end{align}
  \begin{align}
    \begin{cases}
      g\rvert_{\mathcal{S}}= \sum_{n=1}^{\infty}\frac{m_{g,n}}{R^n}\\
      b\rvert_{\mathcal{S}}= \sum_{n=1}^\infty \frac{m_{b,n}}{R^n}\\
      u\rvert_{\mathcal{S}}= \sum_{n=1}^\infty \frac{m_{u,n}}{R^n}
    \end{cases}\,,
    \begin{cases}
      \nabla_Tg\rvert_{\mathcal{S}}= O_{\mathcal{S}}(R^{-2})\\
      \nabla_Tb\rvert_{\mathcal{S}}= O_{\mathcal{S}}(R^{-2})\\
      \nabla_Tu\rvert_{\mathcal{S}}= O_{\mathcal{S}}(R^{-2})
    \end{cases}\,,
  \end{align}
  where~$m_{\phi,n}$ are scalar functions that are independent of~$T$
  and~$R$. This is valid outside a compact ball centered at~$R=0$.
\end{thm}

\begin{remark} As can be seen from~\eqref{g1b1u1}, a sufficient condition
  to make the~$\log R$ term in~$U_2$ vanish, is that,
\begin{align}
	\cancel{\Delta}m_{u,1} = 0\,,
\end{align}
which is exactly the same requirement as in the flat case. Although,
with a general metric we cannot expect the~$\log R$ terms to vanish at
all orders, because order~$n=3$ in~$u$ is already coupled to the~$b$
equation and might therefore inherit up to one power of~$\log R$,
depending on the form of the metric functions.
\end{remark}
\begin{remark}
Once again one can see that the \textit{good-bad-ugly} system admits a
static solution given by,
\begin{align}
    &Rg=m_{g,1}\,,\non\\
    &Rb=m_{b,1}\,,\non\\
    &Ru=m_{u,1}\,,
\end{align}
as long as the requirement~$\cancel{\Delta}m_{\phi,1}=0$ is
fulfilled. If all further initial data are set to zero, then the
series truncates at~$n=1$ and all orders vanish except the first
one. Note that this static solution differs from~\eqref{static1}
because here the operator~$\cancel{\Delta}$ is not necessarily the
Laplace operator on the 2-sphere of radius $R$, but an analogous
operator constructed from~$\slashed{g}^{ab}$ that coincides with the
former to leading order.
\end{remark}

\section{Stratified null forms} \label{section:nullforms}

We can generalize this proof to encompass models more complicated than
the standard \textit{good-bad-ugly system}. In fact there is a large
class of terms that, added to the RHS of~\eqref{gbu} require no
significant changes in the induction proof. These terms are a
generalization of the classical null forms, see~\cite{Kla80,Sog95},
that know about the different types of field. Let us define
\textit{stratified null forms} as terms that involve up to one
derivative of the evolved fields and fall-off faster than~$R^{-2}$
close to null infinity. For example, a term which has one good
derivative, one bad derivative and no explicit dependence on
coordinates, say,
\begin{align}
	\nabla_{\ul{\psi}}g\nabla_\psi b\,,
\end{align}
is necessarily~$o^+(R^{-2})$, and is therefore a stratified null
form. Another type of term that fulfills this requirement is one which
is quadratic in bad derivatives, but has one power of~$R^{-1}$, say,
\begin{align}
	\frac{1}{R}\nabla_{\ul{\psi}}g\nabla_{\ul{\psi}} b\,.
\end{align}
Finally, a term where any derivative hits an ugly field and a bad
derivative hits any field, say,
\begin{align}
	\nabla_au\nabla_{\ul{\psi}} b\,,
\end{align}
is also a stratified null form, because any derivative hitting an ugly
field, necessarily improves its decay. We will need this definition
because it distinguishes the terms that significantly change our proof
from those that do not. Let us replace our earlier system with
\begin{align}\label{gbuN}
   &\mathring{\square} g = N_g\,,\non\\
   &\mathring{\square} b = (\nabla_T g)^2 + N_b\,,\non\\
   &\mathring{\square} u = \tfrac{2}{R}\mn_T u + N_u\,,
\end{align}
where~$N_\phi$ are arbitrary linear combinations of stratified null
forms. As stratified null forms are at least of order~$o^+(R^{-2})$,
regardless of any of these terms we add to the RHS of the
\textit{good-bad-ugly} system, the first order
equations~\eqref{g1b1u1} remain the same, as they are the result of
collecting terms proportional to~$R^{-2}$ ($R^{-3}$ in the~$u$ case),
\begin{align}
    &\nabla_{\psi}\nabla_T\mathcal{G}_1\simeq 0\,,\non\\
    &2R\nabla_{\psi}\nabla_T\mathcal{B}_1\simeq
    -(\nabla_TG_1)^2\,,\non\\
    &2R\nabla_{\psi}\nabla_T\mathcal{U}_2 \simeq
    \tilde{\cancel{\Delta}}m_{u,1}\,.
\end{align}
Stratified null forms will, in general, contribute to the next order,
however they will not contain derivatives of~$\mathcal{G}_2$,
$\mathcal{B}_2$ or~$\mathcal{U}_3$. This is true for all~$n$. At each
step we collect terms of order~$R^{-n-1}$ to find equations
for~$\mathcal{G}_n$, $\mathcal{B}_n$ or~$\mathcal{U}_n$ and stratified
null forms will only contain derivatives
of~$\{G_{m,k},B_{m,k},U_{m,k},m_{u,1}\}$, for~$m\in[1,n-1]$
and~$k\in[0,m]$. This implies that any terms arising from stratified
null forms can be absorbed into~$\Omega^{\phi}_{n-1}$ so that we get
(cf. \eqref{recur1}),
\begin{align}
  &R^{n}\nabla_{\psi}\left(\frac{1}{R^{n-1}}
        \nabla_T\mathcal{G}_n\right) \simeq \Omega'^g_{n-1}\,,\non\\
  &R^{n}\nabla_{\psi}\left(\frac{1}{R^{n-1}}
        \nabla_T\mathcal{B}_n\right) + \nabla_TG_n
        \nabla_TG_1 \simeq \Omega'^b_{n-1}\,,\non\\
  &R^{n-1}\nabla_{\psi}\left(\frac{1}{R^{n-2}}
        \nabla_T\mathcal{U}_n\right) \simeq \Omega'^u_{n-1}\,,
\end{align}
where~$\Omega'^\phi_{n-1}$ are just~$\Omega^\phi_{n-1}$, as defined
earlier, plus any extra terms coming
from~$N_\phi$. Naturally,~\eqref{expH} is still valid and hence~$g$,
$b$ and~$u$ can be written as polyhomogeneous
functions~\eqref{ggeneral},~\eqref{bgeneral}
and~\eqref{ugeneral}. Although we can expect the final recursion
relations of the~$g$, $b$ and~$u$ fields to change in general, the
induction proof remains unchanged. Therefore, regardless of the
addition of any stratified null forms to the \textit{good-bad-ugly}
system, we have the following result,
\begin{align}
  &g = \sum_{n=1}^{\infty}\sum_{k=0}^{n-1}
  \frac{(\log R)^k G_{n,k}(\psi^*)}{R^n}\non\\
  &b =  \sum_{n=1}^{\infty}\sum_{k=0}^{n}
  \frac{(\log R)^k B_{n,k}(\psi^*)}{R^n}\\
  &u =  \frac{m_{u,1}}{R}+\sum_{n=2}^{\infty}
  \sum_{k=0}^{n-1} \frac{(\log R)^k U_{n,k}(\psi^*)}{R^n}\non\,,
\end{align}
This implies that we can generalize \textit{Theorem} \ref{T2} in order
to incorporate stratified null forms.

\begin{thm} \label{T3}
  Let $X^{\ul{\alpha}}=(T,X^{\ul{i}})$ be an asymptotically
  Cartesian coordinate system with an associated covariant
  derivative $\mn$. The \textit{good-bad-ugly} system defined
  as,
\begin{align}
    &\mathring{\square} g = N_g\,,\non\\
    &\mathring{\square} b = (\nabla_T g)^2 + N_b\,,\non\\
    &\mathring{\square} u = \tfrac{2}{R}\mn_T u + N_u\,,
\end{align}
where~$N_\phi$ are arbitrary linear combinations of stratified null
forms, $\mathring{\square}:=g^{ab}\mn_a\mn_b$ and~$g_{ab}$ is an
asymptotically flat metric, admits a polyhomogeneous expansion near
null infinity of the type,
\begin{align}
  &g = \sum_{n=1}^{\infty}\sum_{k=0}^{n-1} \frac{(\log R)^k
    G_{n,k}(\psi^*)}{R^n}\non\\
  &b =  \sum_{n=1}^{\infty}\sum_{k=0}^{n} \frac{(\log R)^k
    B_{n,k}(\psi^*)}{R^n}\non\\
  &u =  \frac{m_{u,1}}{R}+\sum_{n=2}^{\infty}\sum_{k=0}^{n-1}
  \frac{(\log R)^k U_{n,k}(\psi^*)}{R^n}\,,
\end{align}
with initial data on~$\mathcal{S}$ of the type,
\begin{align}
  \begin{cases}
    g\rvert_{\mathcal{S}}= \sum_{n=1}^{\infty}\frac{m_{g,n}}{R^n}\\
    b\rvert_{\mathcal{S}}= \sum_{n=1}^\infty \frac{m_{b,n}}{R^n}\\
    u\rvert_{\mathcal{S}}= \sum_{n=1}^\infty \frac{m_{u,n}}{R^n}
  \end{cases}\,,
  \begin{cases}
    \nabla_Tg\rvert_{\mathcal{S}}= O_{\mathcal{S}}(R^{-2})\\
    \nabla_Tb\rvert_{\mathcal{S}}= O_{\mathcal{S}}(R^{-2})\\
    \nabla_Tu\rvert_{\mathcal{S}}= O_{\mathcal{S}}(R^{-2})
  \end{cases}\,,
\end{align}
where~$m_{\phi,n}$ are scalar functions that are independent of~$T$
and~$R$. This is valid outside a compact ball centered at~$R=0$.
\end{thm}

\section{Conclusions}\label{conclusion}

In this paper we laid out a heuristic method to predict the decay of
terms beyond first order in~$R^{-1}$ in the \textit{good-bad-ugly}
system. In its most general form the model consists of a set of
coupled nonlinear wave equations in which the three different classes
of fields have different asymptotic properties near null-infinity. We
began with the simplest form of this system, as introduced
in~\cite{GasGauHil19}, built from the Minkowski metric and found that
near null infinity there exist formal solutions to this model in which
the bad field may have~$\log R$ terms at every order in~$R^{-1}$, the
ugly field may have logs from second order onward, whereas the good
field has no logs at all. We showed furthermore a recursion relation
that allows us to find each order in~$R^{-1}$ from the previous one to
arbitrary order. The method is, however, heuristic because we have not
shown that all physically relevant solutions of the
\textit{good-bad-ugly} system admit expansions of this form.

This was used as a warm-up for a more general system built from a
general asymptotically flat metric. Keeping the metric functions
fairly general, insisting essentially only that they be analytic
functions of the evolved fields, we showed by induction that there is
a class of asymptotic solutions near null infinity characterized by
polyhomogeneous functions, the main difference between the three types
of fields being the order at which log terms are first allowed to
appear. As the metric components were intentionally left free, a full
recursion relation for a general metric was not possible. However, we
anticipate no reason why this method would not be straightforwardly
applicable to any metric with these requirements in order to find such
relations. In a final generalization to the model we considered the
effect of non-linearities of a special class that we call stratified
null forms. By definition these are precisely the terms involving up
to one derivative of the evolved fields that fall-off faster
than~$O(R^{-2})$. All of our results are subsumed within
Theorem~\ref{T3}, which says that the same type of expansion also
works out in the presence of arbitrary stratified null forms.

The restriction of having just one field of each type in our model is
purely for simplicity. A more general setup with sets of fields of
each type just requires more book-keeping. In fact, in future work, we
aim to apply this method to the EFE in GHG to predict that its
asymptotic solutions can be written as polyhomogeneous functions near
null infinity. Due to the complexity of the full field equations, the
asymptotic system will presumably be very long, but we anticipate that
the the non-linearities studied in the \textit{good-bad-ugly} system
already capture the subtleties of those in GR. By finding the first
few orders of a polyhomogeneous expansion of asymptotic solutions to
the EFE, we expect to be able to recover the peeling properties, or a
polyhomogeneous generalization, of the gravitational field.

Similar polyhomogeneous behavior of the gravitational field close to
spatial and null infinity has been obtained by means of the conformal
Einstein field equations in~\cite{GasVal18, Fri98a} ---see
also~\cite{ChrMacSin95, NovGol82, Win85, KlaNic03} for further
discussion of peeling. The polyhomogeneous expansions described
in~\cite{GasVal18, Fri98a} are formal in the sense that the
appropriate energy estimates needed to rigorously prove that these
expansions arise as an actual solution from some given initial data
are still lacking. The polyhomogeneity result we potentially expect to
obtain by exploiting the methods presented above for the EFE in GHG
would be formal in the same sense. Ultimately we aim to make contact
with the expansions given in~\cite{SatWal19} in harmonic gauge, in
which no log terms are present. It is worth mentioning that the
logarithmic terms appearing in the expansions described
in~\cite{GasVal18, Fri98a} have a very different origin from those
analyzed here. In the case of the \textit{good-bad-ugly} model and the
EFE in GHG the logarithmic terms appear in the asymptotic expansion
due to the form of the non-linearities in the equations, while the
logarithmic terms of~\cite{GasVal18, Fri98a} appear even in a linear
context, such as the spin-2 field equations, in a Minkowski background
in the framework of the cylinder at spatial infinity as discussed
in~\cite{Val03a}.

We furthermore hope that this work will be a stepping stone towards a
full regularization of GR in GHG at null-infinity. Knowing from the
outset where the logs may appear up to arbitrary order, one can employ
a `subtract-the-logs' strategy as the one used in~\cite{GasGauHil19}
in order to treat these divergent terms, or indeed attempt to
carefully choose gauge source functions that eradicate them all
together.

\acknowledgments

The Authors wish to thank Alex Va\~{n}\'o-Vi\~{n}uales for helpful
comments on the manuscript. MD acknowledges support from FCT
(Portugal) program PD/BD/135511/2018, DH acknowledges support from the
FCT (Portugal) IF Program IF/00577/2015, PTDC/MAT- APL/30043/2017. JF
acknowledges support from FCT (Portugal) programs PTDC/MAT-APL/30043/2017,
UIDB/00099/2020. EG gratefully acknowledges support from the European
Union’s H2020 ERC Consolidator Grant “Matter and Strong-Field Gravity:
New Frontiers in Einstein’s Theory,” Grant Agreement No. MaGRaTh-646597.
EG also acknowledges support from the European Union (through the PO
FEDER-FSE Bourgogne 2014/2020 program) and the EIPHI Graduate School
(contract ANR-17-EURE-0002) as part of the ISA 2019 project.

\normalem
\bibliography{asymexp}

\end{document}